\documentclass[12pt]{article}
\usepackage{graphicx}
\usepackage{axodraw}
\begin{document}


\def\a{\alpha}
\def\b{\beta}
\def\c{\varepsilon}
\def\d{\delta}
\def\e{\epsilon}
\def\f{\phi}
\def\g{\gamma}
\def\h{\theta}
\def\k{\kappa}
\def\l{\lambda}
\def\m{\mu}
\def\n{\nu}
\def\p{\psi}
\def\q{\partial}
\def\r{\rho}
\def\s{\sigma}
\def\t{\tau}
\def\u{\upsilon}
\def\v{\varphi}
\def\w{\omega}
\def\x{\xi}
\def\y{\eta}
\def\z{\zeta}
\def\D{\Delta}
\def\G{\Gamma}
\def\H{\Theta}
\def\L{\Lambda}
\def\F{\Phi}
\def\P{\Psi}
\def\S{\Sigma}

\def\o{\over}
\def\beq{\begin{eqnarray}}
\def\eeq{\end{eqnarray}}
\newcommand{\gsim}{ \mathop{}_{\textstyle \sim}^{\textstyle >} }
\newcommand{\lsim}{ \mathop{}_{\textstyle \sim}^{\textstyle <} }
\newcommand{\vev}[1]{ \left\langle {#1} \right\rangle }
\newcommand{\bra}[1]{ \langle {#1} | }
\newcommand{\ket}[1]{ | {#1} \rangle }
\newcommand{\EV}{ {\rm eV} }
\newcommand{\KEV}{ {\rm keV} }
\newcommand{\MEV}{ {\rm MeV} }
\newcommand{\GEV}{ {\rm GeV} }
\newcommand{\TEV}{ {\rm TeV} }
\def\diag{\mathop{\rm diag}\nolimits}
\def\Spin{\mathop{\rm Spin}}
\def\SO{\mathop{\rm SO}}
\def\O{\mathop{\rm O}}
\def\SU{\mathop{\rm SU}}
\def\U{\mathop{\rm U}}
\def\Sp{\mathop{\rm Sp}}
\def\SL{\mathop{\rm SL}}
\def\tr{\mathop{\rm tr}}

\def\IJMP{Int.~J.~Mod.~Phys. }
\def\MPL{Mod.~Phys.~Lett. }
\def\NP{Nucl.~Phys. }
\def\PL{Phys.~Lett. }
\def\PR{Phys.~Rev. }
\def\PRL{Phys.~Rev.~Lett. }
\def\PTP{Prog.~Theor.~Phys. }
\def\ZP{Z.~Phys. }


\baselineskip 0.7cm

\begin{titlepage}

\begin{flushright}
UT-03-30
\end{flushright}

\vskip 1.35cm
\begin{center}
{\large \bf
Thermal Leptogenesis and Gauge Mediation
}
\vskip 1.2cm
M.~Fujii${}^{1}$, M.~Ibe${}^{1}$ and T.~Yanagida${}^{1,2}$
\vskip 0.4cm

${}^1${\it Department of Physics, University of Tokyo,\\
     Tokyo 113-0033, Japan}

${}^2${\it Research Center for the Early Universe, University of Tokyo,\\
     Tokyo 113-0033, Japan}

\vskip 1.5cm

\abstract{
We show that a mini-thermal inflation occurs naturally in a class of
gauge mediation models of supersymmetry (SUSY) breaking, provided that 
the reheating temperature $T_R$ of the primary inflation is much higher 
than the SUSY-breaking scale, say $T_R > 10^{10}$ GeV. The reheating
process of the thermal inflation produces an amount of entropy, which 
dilutes the number density of relic gravitinos. This dilution renders 
the gravitino to be the dark matter in the present universe. The
abundance of the gravitinos is independent of the reheating temperature
 $T_R$, once the gravtinos are thermally  produced after the reheating
 of the primary inflation. 
We find that the thermal leptogenesis  takes place at $T_L\simeq 10^{12-14}$  
GeV for $m_{3/2}\simeq 100$ keV$-10$ MeV without any
gravitino problem. 
}
\end{center}
\end{titlepage}

\setcounter{page}{2}

\section{Introduction}

The baryon-number asymmetry in the universe is one of the fundamental
parameters in cosmology. 
There have been proposed a number of mechanisms for producing the baryon
asymmetry in the early universe. 
Among them the leptogenesis \cite{Fukugita:1986hr} is the most
attractive and fruitful mechanism, since it may have a connection to the
low-energy observation, that is, neutrino masses and mixings. 
In fact, a detailed analysis on the thermal leptogenesis \cite{BBP}
gives an upper bound on all neutrino masses of 0.1 eV, which is
consistent with data of neutrino oscillation experiments.  
The thermal leptogenesis requires the reheating temperature $T_R \gsim 
10^{10}$ GeV, which, however, leads to overproduction of unstable
gravitinos.
The decays of gravitinos produced after inflation destroy the success of
nucleosynthesis \cite{Weinberg:zq,Holtmann:1998gd}.
This problem is not solved even if one raises the gravitino mass
$m_{3/2}$ up to 30 TeV \cite{Kohri:2001jx} and hence the thermal
leptogenesis seems to have a tension with the gravity mediation model of
supersymmetry (SUSY) breaking.  
In the gauge mediation model \cite{gauge-mediation}, on the 
other hand, the gravitino is the lightest SUSY particle (LSP) and one
does not need to worry about the gravitino decay. 
However, we have even a stronger constraint on the reheating temperature
$T_R \lsim 10^{8}$ GeV for $m_{3/2} \lsim 1$ GeV to avoid the
overproduction of the gravitinos \cite{Moroi:1993mb} (otherwise, the stable
gravitinos overclose the present universe).  

It has been recently pointed out \cite{Fujii:2002fv} that the above problem is
naturally solved in the gauge mediation model. The crucial observation
is that the gravitinos are  in the thermal equilibrium at high
temperatures such as $T_R \gsim 10^{10}$ GeV for $m_{3/2} \lsim 1$ GeV.
Thus, the number density of gravitinos is independent of the reheating
temperature once they are in thermal equilibrium, while the maximal lepton 
(baryon) asymmetry depends linearly on $T_R$. Therefore, if a suitable
amount of entropy is provided at later time to dilute the number density
of gravitinos, one may account for both the dark matter abundance
and the baryon asymmetry in the present universe.
More interestingly, Ref.\cite{Fujii:2002fv} shows that the required entropy is
naturally supplied by decays of messenger particles in the gauge mediation
model if the SUSY-breaking transmission to the SUSY standard-model (SSM)
sector is direct. In this paper we show that the late-time entropy
production takes place even if the mediation of SUSY breaking is NOT of
the direct type. This is because the gauge mediation model we discuss
in this paper has a flat potential and a mini-thermal inflation
occurs naturally producing the required amount of entropy.

\section{The gauge mediation model}\label{sec:model}

We consider an extension of the gauge mediation model proposed in
\cite{Nomura:1997ur}. The reason why we take this model is that it has the
unique true vacuum of SUSY breaking. Otherwise, it seems very difficult
to choose a SUSY-breaking false vacuum in the evolution of the universe,
since we assume the reheating temperature $T_R$ much higher than the
SUSY-breaking scale.

The dynamical SUSY-breaking (DSB) sector is based on a SUSY SU(2)$_H$
hypercolor gauge theory with four doublet chiral superfields $Q^i$
called hyperquarks and six singlet ones $Z^{ij} = -Z^{ji}$
\cite{Izawa:1996pk,Intriligator:1996pu}.  
Here, the indices $\alpha = 1, 2$ denote the SU(2)$_H$ ones and the 
indices $i,j=1,...,4$ are flavor ones.  
We impose, for simplicity, a flavor symmetry SP(4) and write the
superpotential as 
\begin{eqnarray}
 W_{\rm tree} = \lambda Z(QQ) + \lambda 'Z^a(QQ)_a,
\label{eq:DSBtree}
\end{eqnarray}
where $Z$ and $(QQ)$ are singlets of the flavor SP(4) and $Z^a$ and
$(QQ)_a$ are ${\bf 5}$ representations of the SP(4).\footnote{
$Z$ and $Z^a$ are linear combinations of the original $Z^{ij}$.
}
It should be noted here that we have a global U(1)$\times$U(1)$_R$ in
addition to the flavor SP(4) at the classical level, where the U(1)$_R$
represents a $R$ symmetry.  
We choose $R$-charges of relevant superfields so that the U(1)$_R$ has
no SU(2)$_{H}$ gauge anomaly.
The $R$-charges for the $Q^i_\alpha$ and $Z_{(a)}$ are given in Table
1. 
The flavor U(1) breaks down to a discrete Z$_4$ symmetry at the quantum
level,  under which $Q^i_\alpha$ transforms as $Q^i_\alpha \rightarrow
iQ^i_\alpha$ and $Z_{(a)}$ as $Z_{(a)}\rightarrow -Z_{(a)}$.

We show that for
$\lambda ' > \lambda$ the low-energy effective superpotential is 
approximately given by
\begin{eqnarray}
 W_{\rm effective} \simeq \frac{\lambda}{(4\pi)^2} \Lambda_H^2Z,
\label{eq:DSBeff}
\end{eqnarray}
where $\Lambda_H$ is a dynamical scale of the SU(2)$_H$ gauge
interaction and  
\begin{eqnarray}
 \langle (QQ)\rangle \equiv \langle \frac{1}{2}(Q_1Q_2 +Q_3Q_4)\rangle \simeq 
\bigg(\frac{\Lambda_H}{4\pi}\bigg)^2.
\label{eq:DSBvev}
\end{eqnarray}
The superfield $Z$ has a non-vanishing $F$ term $\langle F_Z\rangle \simeq
\lambda \Lambda_H ^2/(4\pi)^2$ and hence SUSY is spontaneously broken 
\cite{Izawa:1996pk,Intriligator:1996pu}.\footnote{
The integration of the hypercolor sector induces a nonminimal Kahler
potential of the $Z$ superfield, which determines the vacuum-expectation
value of the $Z$ field.
However, one can not calculate the Kahler potential due to the strong
hypercolor gauge interatction.
Thus, we postulate $\vev{Z}=0$, for simplicity.
}
The condensation of $QQ$ does not break the $R$ symmetry, but causes the
breaking of the discrete Z$_4$ symmetry down to a discrete Z$_2$. 
This breaking generates unwanted domain walls and hence we should
introduce explicit breaking terms of the Z$_4$. 
Here, we introduce a nonrenormalizable interaction in the Kahler
potential, $K=(k/M_G^2)QQZZ^{\dagger}$, to eliminate the domain walls
before they dominate the early universe,\footnote{
We find that this breaking term with $k={\cal O}(1)$ is strong enough to
eliminate the unwanted domain walls (see \cite{Vilenkin:zs}).
}
where $M_G$ is the gravitational scale $M_G\simeq 2.4 \times 10^{18}$
GeV.  
We see that this nonrenormalizable interaction does not affect the
dynamics of SUSY breaking. 

\begin{figure}[htb]
\begin{center}
\includegraphics[width=0.35\linewidth,height=0.2\linewidth]{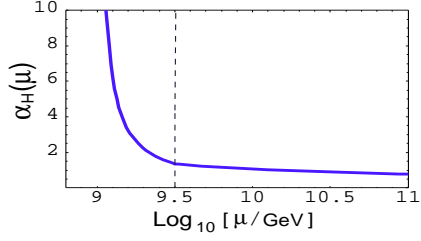}
\caption{The running of gauge coupling constant $\alpha_H$. 
 $\mu$ denotes the renormalization point.
Here, we assume $n_q = 3$, $M_{Q'} = 10^{9.5}$ GeV and $\alpha_H(M_G) = 0.25$. 
The vertical dashed line denotes the mass scale $M_{Q'}$.
}
\label{fig:alpha}
\end{center}
\end{figure}

We now introduce 2$n_q$ massive hypercolor quarks $Q'^{a(j)}_{\alpha}$
($j=1,\cdots,n_q$) and assume that each pairs of the
$Q'^{a(j)}_{\alpha}$ form doublets ($a=1,2$) of a new gauge group
SU(2)$_m$.  
Namely, the massive hyperquarks are (${\bf 2,2}$) representations of
SU(2)$_H\times$SU(2)$_m$.
In the original model in \cite{Nomura:1997ur} a U(1)$_m$ subgroup of the
SP(4) is  gauged. 
The reason why we introduce the SU(2)$_m$ gauge interaction becomes
clear in the next section.
Notice that the introduction of the above massive hyperquarks does not
affect the dynamics of the SUSY breaking \cite{Izawa:1997hu}.
The SU(2)$_m$ gauge interaction and the massive hyperquarks
$Q'^{a(j)}_\alpha$ play a role of transmitting the SUSY 
breaking effects to the messenger sector. 
In the present analysis we take the masses $M_Q'$ of the hyperquarks 
$Q'^{a(j)}_{\alpha}$ at the dynamical scale of the hypercolor SU(2)$_H$
gauge interaction, that is $M_{Q'} \simeq \Lambda_H$.
As we see from Fig. \ref{fig:alpha}, this assumption is
natural since the running of the gauge coupling constant $\alpha_H$
becomes very fast below the mass scale $M_{Q'}$. 

The messenger sector consists of $2n$ chiral superfields
$E_{a}^{i}$ with $i=1,..., 2n$
which are doublets of the SU(2)$_m$, a singlet superfields $S$, and
vector-like messenger quark and lepton superfields, $d_M,~{\bar d}_M,~\ell_M$
and ${\bar \ell}_M$.\footnote{
The messenger fields, ($\bar{d}_M$, $\ell_M$) and ($d_M$,
$\bar{\ell}_M$)  transform as {\bf 5}$^*$ and {\bf 5} of the grand
unification group SU(5)$_{\rm GUT}$, respectively.
}

For $n\leq 2$, the SU(2)$_m$ symmetry is broken after the DSB  sector is
integrated out. 
Thus, we take $n=3$ in the present analysis. 
Other cases will be discussed elsewhere.\footnote{
For $n = 5$ one may assign $E^i_{\alpha}$ ($i=1\sim 10$) to be {\bf
5}$+${\bf 5}$^*$ of the SU(5)$_{\rm GUT}$.
In this case one does not need to introduce the messenger quarks and
leptons.
}
The most general superpotential for the messenger sector without any
dimensional parameters is 
\begin{eqnarray}
 W_{\rm mess} = \sum_{i\neq j} k_{ij}SE^{i}E^{j} + \frac{f}{3}S^3 + k_d S d_M {\bar d}_M + 
k_\ell S\ell_M{\bar \ell}_M,
\label{eq:messengerpotential}
\end{eqnarray}
where $k_{ij}=-k_{ji}$ ($i,~j=1,..., 6$), and we have omitted 
indices of the messenger gauge SU(2)$_m$. 
This $W_{\rm mess}$ is natural, since we have a $R$ symmetry that
forbids other possible terms in the superpotential. 
$R$ charges for relevant superfields are given in Table
\ref{tab:R-charge}. 
It is clear that the superpotential $W_{\rm mess}$ possesses a discrete
Z$_3$ symmetry where $S,E,d_M$ and $\ell_M$ have the Z$_3$ charge $+1$.

\begin{table}[t]
\begin{center}
\begin{tabular}{c|cccccccc}
Fields   & $Z^{ij}$ &  $Q_i$ & $ S $ & $E^i$ & $d_M$, $\bar{\ell}_M$ & $ \bar{d}_M$, 
$\ell_M$ & $\bar{d}$ & $\ell$  \\ 
\hline 
$R$ charges & 2 & 0  & 2/3 & 2/3 & -1 & 7/3 & 1 & 1  \\
\end{tabular}
\caption{$R$ charges of the fields in the DSB, the messenger and the 
SSM sectors.}
\label{tab:R-charge}
\end{center}
\end{table}

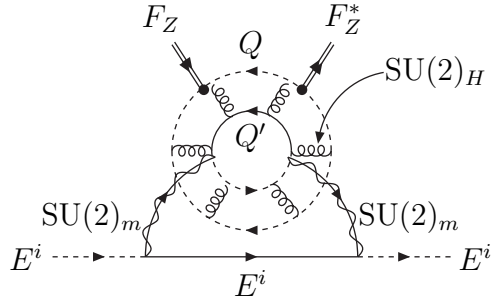
\begin{figure}[htbp]
\begin{center}
  \begin{picture}(150,100)(0,0)
  \DashArrowLine(0,0)(35,0){2}
  \ArrowLine(35,0)(115,0)
  \DashArrowLine(115,0)(150,0){2}
  \PhotonArc(75,0)(40,110,180){2}{4}
  \ArrowArcn(75,0)(40,180,110)
  \PhotonArc(75,0)(40,0,70){2}{4}
  \ArrowArcn(75,0)(40,70,0)
  \ArrowArc(75,40)(15,0,180)
  \DashArrowArc(75,40)(15,180,0){2}
  \DashArrowArc(75,40)(30,0,180){2}
  \DashArrowArcn(75,40)(30,0,180){2}
  \ArrowLine(45,80)(56,63)
  \ArrowLine(46,81)(57,64)
  \ArrowLine(94,64)(105,81)
  \ArrowLine(95,63)(106,80)
  \Vertex(94,63){2}
  \Vertex(57,63){2}
  \Gluon(65,27)(58,15){2}{3}
  \Gluon(82,27)(92,15){2}{3}
  \Gluon(68,53)(61,65){2}{3}
  \Gluon(82,53)(89,65){2}{3}
  \Gluon(45,40)(60,40){2}{3}
  \Gluon(105,40)(90,40){2}{3}
  \Text(75,80)[]{$Q$}
  \Text(75,45)[]{$Q'$}
  \Text(-10,0)[]{${E}^{i}$}
  \Text(160,0)[]{${E}^{i}$}
  \Text(75,-10)[]{${E}^{i}$}
  \Text(135,15)[]{SU(2)$_m$}
  \Text(15,15)[]{SU(2)$_m$}
  \Text(110,90)[]{$F_Z^*$}
  \Text(41,90)[]{$F_Z$}
  \Text(145,70)[]{SU(2)$_H$}
  \LongArrowArc(130,40)(30,100,175)
 \end{picture}
\end{center}
\caption{A typical example of the Feynman diagrams which give the soft
 SUSY-breaking masses for $E^i$. } 
\label{fig:softE}
\end{figure}

The SU(2)$_m$ gauge interaction (together with the hypercolor SU(2)$_H$
interaction) transmits the SUSY-breaking effects of the DSB sector to
the messenger sector and generates soft SUSY-breaking masses for the
$E_\alpha^i$ superfields, $(m^{soft}_E)^2$ (for an example of Feynman 
diagrams see Fig. \ref{fig:softE}).   
A straitforward calculation \cite{Nomura:1997ur} shows (see also
\cite{Izawa:1997hu}) 
\begin{eqnarray}
 m^{soft}_E \simeq \sqrt{\frac{3n_q}{2}}\bigg(\frac{\alpha_m}{4\pi}\bigg) 
\frac{\lambda F_Z}{\Lambda_H}\simeq\sqrt{\frac{3n_q}{2}} 
\bigg(\frac{\alpha_m}{4\pi}\bigg)\frac{\lambda ^2}{16\pi ^2}         
\Lambda_H,  
\label{eq:softmE}
\end{eqnarray}
where $\alpha_m =g^2_m/4\pi$ with $g_m$ being the SU(2)$_m$ gauge
coupling constant. As shown in Ref.\cite{Nomura:1997ur}, effects of
$E^i$ loops give rise to a negative soft SUSY-breaking mass squared,
$-m^2_S$, for the singlet superfield $S$.  
The $m^2_S$ is estimated as
\begin{eqnarray}
  m^2_S \simeq \frac{16}{16\pi^2}\sum_{i\neq j} k^2_{ij}(m^{soft}_E)^2
               \ln\frac{\Lambda_H}{M_E}, 
\label{softmS}
\end{eqnarray}
where $M_E$ is a SUSY-invariant mass for the superfields $E^i$ which is
given by the condensation of the superfield $S$
(see the following discussion).  

Now we have a potential for the scalar fields in the messenger sector,  
\begin{eqnarray}
\nonumber
  V_{\rm mesenger} 
&\simeq& \big|\sum_{i\neq j}k_{ij}E^iE^j + fS^2 + k_dd_M{\bar d}_M 
                          + k_\ell\ell_M{\bar \ell}_M\big|^2 
                          +\sum_{i=1}^{6} |\sum_{j\neq i} k_{i,j} S E^j|^2 \\ \nonumber
                        &  &+ |k_d S d_M|^2 + |k_d S \bar{d}_M|^2
                           + |k_l S \ell_M|^2 + |k_l S \bar{\ell}_M|^2\\ 
			& &  + (m^{soft}_E)^2 \sum^6_{i=1}|E^i|^2
                          -m_S^2|S|^2, 
\label{eq:effpotential}
\end{eqnarray}
where all fields represent corresponding scalar boson fields. We
see that this potential has a global minimum at
\begin{eqnarray}
  \langle S^*S\rangle = \frac{m_S^2}{2f^2}, \quad
  \langle E^i \rangle = \langle d_M\rangle = \langle\bar{d}_M\rangle 
= \langle \ell_M\rangle = \langle\bar{\ell}_M\rangle = 0, \quad 
 \vev{|F_S|} = \frac{m_S^2}{2f},
\label{eq:messengervev}
\end{eqnarray}
for $k_d,~k_\ell \gg f$.
We show in section \ref{sec:thermalif} that this condition for the
Yukawa coupling constants is naturally realized.   
The superfields $E^i$, $d_M$ $(\bar{d}_M)$ and $l_M$ $(\bar{l}_M)$ have
SUSY-invariant masses as $M_E = 2 k_{ij} \vev{S}$, $M_d =  k_d \vev{S}$
and $ M_l =  k_l \vev{S}$, respectively.  
The SUSY-breaking effects are transmitted to the messenger quark and
lepton superfields through $\vev{F_S}$ and Yukawa coupling in Eq.
(\ref{eq:messengerpotential}).

The condensation of the $S$ field, $\vev{S}\neq 0$, breaks the $R$
symmetry which generates a $R$ axion (the phase component of the complex
$S$ boson).  
The axion mass is usually induced by a constant term in the
superpotential, since the constant term breaks the $R$ symmetry
explicitly. 
However, in the present model the induced axion mass vanishes at the
tree level , and hence we need another explicit breaking term of the $R$
symmetry to give a sufficiently large mass to the $R$ axion. 
We introduce a nonrenormalizable interaction in the superpotential,
$W_{\rm mess} = (1/M_G)QQS^2$.\footnote{
We assume, throughout this paper, that the $R$ symmetry is explicitly
broken by nonrenormalizable interactions suppressed by the gravitational
scale. 
}  
This new term induces the $R$ axion mass as $m_{\rm  axion}\simeq 10
\GEV\sqrt{(m_{3/2}/\MEV)(m_S/10^5\GEV)}$.  
Notice that this nonrenormalizable interaction breaks also the discrete
Z$_3$ symmetry explicitly and hence we have no domain-wall problem.  

The SSM gauginos acquire soft SUSY-breaking masses through messenger
loop diagrams, and at the one-loop level they can be written as
\cite{gauge-mediation}
\begin{eqnarray}
 m_{\tilde{g}_i} = c_i \frac{\alpha_i}{4\pi} M_{\rm mess},
\label{eq:gaugino}
 \end{eqnarray}
where $c_1 = 5/3$, $c_2 = c_3 = 1$, and $m_{\tilde{g}_i} (i=1,2,3)$
denote the bino, wino and gluino masses, respectively.
Similarly, the soft SUSY-breaking masses for the squraks, sleptons, and
Higgs bosons, $\tilde{f}$, in the SSM sector are generated at the
two-loop level as \cite{gauge-mediation}
\begin{eqnarray}
 m_{\tilde{f}}^2 = 2 M_{\rm mess}^2 \bigg[ C_3 \bigg(\frac{\alpha_3}{4\pi}\bigg)^2
+ C_2 \bigg(\frac{\alpha_2}{4\pi}\bigg)^2 + \frac{5}{3}Y^2 \bigg(\frac{\alpha_1}{4\pi}
\bigg)^2 \bigg],
\label{eq:sfermion}
\end{eqnarray}
where $C_3=4/3$ for color triplets and zero for singlets, $C_2=3/4$ for
weak doublets and zero for singlets,  and $Y$ is the SM hypercharge, $Y
= Q_{\rm em} - T_3$.
Here, $M_{\rm mess}$ is an effective messenger scale defined as
\begin{eqnarray}
 M_{\rm mess} \equiv \frac{\vev{|F_S|}}{\vev{S}} = \frac{m_S}{\sqrt{2}},
\label{eq:lambdamess1}
\end{eqnarray}
and in terms of SUSY-breaking scale  $\sqrt{F_Z}$, it can be written as 
\begin{eqnarray}
 M_{\rm mess}\simeq \frac{2\sqrt{3n_q}}{(4\pi)^3}\alpha_m \sqrt{\sum_{i\neq j}k_{ij}^2  
\lambda^3\ln\frac{\Lambda_H}{M_E}} \sqrt{F_Z}.
\label{eq:lambdamess2}
\end{eqnarray}
To have the SSM gaugino and sfermion masses at the electroweak scale,
the effective messenger scale $M_{\rm mess}$ must be $\sim
10^{4-5}$ GeV.  
Then, the SUSY-breaking scale $\sqrt{F_Z}$ becomes $\simeq 10^{7-8}$
GeV for $\alpha_m(\Lambda_H) = 0.5$, $\lambda= \sqrt{\sum_{i\neq
j}k_{ij}^2} = {\cal O}(1)$, and $\sqrt{\ln\Lambda_H/M_E} = {\cal
O}(1)$.\footnote{ 
The SU(2)$_m$ gauge coupling constant at $\mu \simeq M_E$ is estimated
as $\alpha_m(M_E)/(4\pi) \simeq 0.2$ for $\alpha_m(\Lambda_H) \simeq
0.5$ and hence the perturbative calculation for the SU(2)$_m$ gauge
interaction at $\mu\simeq M_E$ is valid. 
} 
This corresponds to the gravitino mass,
\begin{eqnarray}
  m_{3/2} \simeq \frac{F_Z}{\sqrt{3}M_G} \simeq 100{\rm keV}-10{\rm MeV}.
\label{eq:gravitino}
\end{eqnarray}
Thus, we consider that the dynamical scale of hypercolor gauge
interaction, $\Lambda_H \simeq 10^{8-9}$ GeV, and the SUSY-breaking
masses for $E_\alpha^i$ and $S$, $m^{soft}_E\simeq 10^{4-5}$ GeV and
$m_S \simeq 10^{4-5}$ GeV.  
We should note here that the SUSY-invariant masses for messenger quarks,
leptons and $E^i$ are about $10^{6-7}$ GeV (see the discussion in
section \ref{sec:thermalif}).   

\section{Decay processes of the quasi-stable states}\label{sec:decay}
We are now at the point to discuss decays of all quasi-stable particles
in the DSB and the messenger sectors and show that their lifetimes are
short enough not to produce extra entropy at the decay times (except for
the $R$ axion).  
We first consider the quasi-stable particles in the DSB sector, that are
the fields $Z^{ij}$, the lightest bound states $Q^iQ^j$, $Q'^iQ'^j$ and 
$Q^iQ'^j$.   

{\bf The DSB sector}\\
The SU(2)$_H$ singlets $Z^{ij}$ may decay into pairs of the SU(2)$_m$
doublets $E^m+E^{l\dagger}$ through  nonrenormalizable interactions 
in the Kahler potential, $K=(h/M_G)Z^{ij}(E^{l} E^{m\dagger}_\alpha)+
h.c.$, with $i,j = 1,\cdots,4 $ and $l,m = 1,..., 6$
(Fig. \ref{fig:DSBdecay}), where $h$ is of order of unity.
The decay rates are estimated as $\Gamma_Z \simeq 6^2(h^2/4\pi)
(hM_E/M_G)^2 M_Z$ and $M_Z \simeq (\lambda^{(')}/4\pi)\Lambda_H$ is a
mass\footnote{
We use a naive dimensional analysis for the hypercolor dynamics
\cite{Luty:1997fk}.
Therefore, we may have a ${\cal O}(1)$ ambiguity in the estimations on
masses and couplings of composite bound states.
}
of the $Z$ and $QQ$.\footnote{ 
The bound states $Q^iQ^j$ form massive multiplets together with the
$Z^{ij}$. 
} 
The decay temperature is $T^{Z}_d \simeq
{\cal O}(100)$ GeV for $M_E \simeq 10^7$ GeV and $\Lambda_H\simeq 10^9$
GeV.\footnote{  
The scalar component of the flat direction $Z$ receives a SUSY-breaking
mass of the order of $\Lambda_H/4\pi$ through one-loop corrections in the
Kahler potential and decays also into a pair of $E+E^{\dagger}$. 
On the contrary, its fermion partner is nothing but the goldstino
component of the gravitino.
}  
The $QQ$ bound states which are mass partners of the $Z^{ij}$ fields
decay similarly.
On the other hand, they decouple from the thermal bath when the rate of
the annihilation $\vev{\sigma v} n_{Z}$ drops below the Hubble
expansion rate $H$, where $\vev{\sigma v}$ is a thermally averaged
annihilation cross section and $n_{Z}$ a number density of $Z^{ij}$
and $QQ$.
Thus, the relic abundance of $Z^{ij}$ and $QQ$ is $n_{Z}\simeq
H/\vev{\sigma v}$ after the decoupling from the thermal bath, and the
total energy density of the universe is given by 
\begin{eqnarray}
 \rho = \frac{\pi^2}{30}g_*(T)T^4 + M_Z n_Z(T), 
\end{eqnarray}
where $g_*(T)$ is the degree of freedoms of effective massless particles
at a temperature $T$. 
Then, if they were stable, they could dominate the energy density of the 
universe ($(\pi^2/30) g_*(T)T^4\lsim M_Z n_Z(T)$) at the
temperature   
\begin{eqnarray}
 T^{Z}_c \simeq \frac{4}{3}M_Z \bigg(\frac{n_Z}{s(T_f)}\bigg)\frac{1}{\Delta_S}
\simeq \frac{M_Z}{\vev{\sigma v}M_G T_f \Delta_S}
\simeq \frac{30M_{Z}^2}{4\pi\alpha_m^2M_G \Delta_S},
\label{eq:Z}
\end{eqnarray}
where $s(T)= (2\pi^2/45)g_*(T)T^3$ is a entropy density, $\Delta_S 
\simeq 10^{2-4}$ the dilution factor\footnote{   
See Eq. (\ref{eq:dilution1}) in the next section.
} 
 from the decay of ``flaton'' $S$,\footnote{
For the definition of the ``flaton'' $S$, see the next section.
}
and $T_f$ their freeze-out temperature. 
We have used $\vev{\sigma v} \simeq 4\pi\alpha_m^2/M_{Z}^2$ and
 $T_f\simeq M_{Z}/30$.\footnote{  
The $QQ$ bound states can annihilate into a pair of the SU(2)$_m$ gauge
multiplets through $Q'Q'$ loop diagrams.
The $Z^{ij}$ which are mass partners of $QQ$ also annihilate into a pair
of the SU(2)$_m$ gauge multiplets through their mass terms.
Thus, annihilation cross section is given by  $\vev{\sigma v}
\simeq (\eta^2/4\pi)(\alpha_m^2/M_{Z}^2)$, where $\eta$ is a factor
which  comes from the strong dynamics and naturally expected to be
${\cal O}(1)$.   
Even if $\eta$ is much smaller than ${\cal O}(1)$ for some reasons, 
the following discussion does not change for $\eta \gsim 0.01$.
}
We can see that $T^Z_c$ is much lower than $T^Z_d$, and hence $Z^{ij}$
and $QQ$ decay before they dominate the universe producing no
significant entropy.  

The SU(2)$_m$ singlet bound states $Q'Q'$ decay into $QQ' + E^i$, and
then the doublet bound states $QQ'$ decay into $QQ + E^i$ through the
Kahler potential $K=(h/M_G)(Q'Q^{\dagger})E^{i\dagger}$
(Fig. \ref{fig:DSBdecay}). 
The decay rates are given by $\Gamma_{\rm hyper} \simeq
(h^2/8\pi) (\Lambda_H/M_G)^2 \Lambda_H$ for both decays and the
corresponding decay temperature is $T_d \simeq {\cal O}(10)$ TeV for
$\Lambda_H\simeq 10^9$ GeV.      
When they were stable, the bound states $Q'Q'$ and $QQ'$ could dominate
the energy density at the temperature $T^{Z}_c$ in Eq. (\ref{eq:Z}),
since their annihilation cross sections and masses are about the same as
those of $QQ$ bound states. 
We can see that $T^{Z}_c$ is much lower than the decay temperatures of
$Q'Q'$ and $QQ'$ and hence they also produce no extra entropy. 

{\bf The messenger sector}\\
We turn to the messenger sector. 
First of all, the lightest $E^iE^j$ bound states can decay into a pair
of $S$ fields through the diagrams in Fig. \ref{fig:decayE}, and decay
rates are given by $\Gamma\simeq (k_{ij}^4/4\pi) M_E$.   
The  SU(2)$_m$ glue balls decay into a pair of $S$ and $S^\dagger$
through one-loop diagrams of intermediate $E_\alpha^{i}$ particles
(Fig. \ref{fig:decayE}), and their decay rates are estimated as $\Gamma
\simeq (6^2/4\pi)(k_E^2\alpha_m/4\pi)^2 (\Lambda_m^3/M_E^2)$.  
We easily see that the decay rates of the $E^iE^j$ bound states and
SU(2)$_m$ glue balls are large enough not to produce extra entropy at
their decay times. 

In the original model in \cite{Nomura:1997ur}, the $E^i$ particles
overclose the universe, since they are stable particles.         
In the present model, however, they have no cosmological problem, since
the $E^i$ particles form necessarily bound states owing to the non-Abelian
gauge dynamics and the bound states can decay sufficiently fast as we
see above.  
This is the main reason why we adopt the non-Abelian SU(2)$_m$ instead of
the U(1)$_m$.

The $S$ fermion and the radial component of the complex $S$ boson (called
$R$ saxion) can decay into a pair of the SSM gauge multiplets through
the messenger quark (lepton) loops, and the decay rate is $\Gamma \simeq
(1/8\pi)(\alpha_3/4\pi)^2 (cm_S/\vev{S})^2 c m_S$ ($c=1/\sqrt{2}$ for $S$
fermion and $c=\sqrt{2}$ for $R$ saxion) (see Fig. \ref{fig:decayS}).      
Thus, the decay rates of those particles are sufficiently large and
hence the $S$ fermion and the $R$ saxion causes no
entropy production.\footnote{ 
As discussed in the next section, a coherent mode of the radial
component of the $S$ boson plays a role of ``flaton'' in the thermal
inflation, which produces a large entropy at the reheating epoch despite
of the large decay rate.}   
In addition to the above decay modes, the $R$ saxions can also decay into
$R$ axion pairs with the decay rate $\Gamma \simeq 
(1/64\pi)(\sqrt{2}m_S/{\vev{S}})^2\sqrt{2}m_S$, which is the dominant
decay mode of the $R$ saxion \cite{Asaka:1999xd}.   
The $R$ axion decays into a QCD gluon pair through the diagram in
Fig. \ref{fig:decayS}, and the decay rate is estimated as $\Gamma \simeq
(1/4\pi)(\alpha_3/4\pi)^2 (m_{\rm axion}/\vev{S})^2 m_{\rm axion}$.  
Thus, their decay temperature is given by 
\begin{eqnarray}
T^{\rm axion}_{d}
\simeq 1\GEV \bigg(\frac{m_{\rm axion}}{10\GEV}\bigg)^{3/2}
\bigg(\frac{10^5 \GEV}{m_S}\bigg)\bigg(\frac{f}{10^{-3}}\bigg).
\end{eqnarray}
As discussed in the next section, the $R$ axion produces a small but
nonnegligible amount of entropy at its decay time and the dilution
factor from the $R$ axion decay is about $10$ (see
Eq. (\ref{eq:dilution2})). 

The remaining stable particles are now messenger quarks and leptons. 
They can mix with the SSM quarks ${\bar d}$ and leptons $\ell$ through
nonrenormalizable interactions, $W= (1/M_G^2)\vev{W} d_M{\bar d} +
(1/M_G^2)\vev{W} \bar{\ell}_M\ell$,  where $\vev{W} \simeq m_{3/2}
M_G^2$ is a constant term in the superpotential which is needed to tune
the vacuum energy vanishing.
The $R$ charges of messenger fields can be chosen so that these
interactions are allowed.
The messenger quarks and leptons decay into the SSM particles through
the mixings \cite{Fujii:2002fv}. 
The decay rate is estimated as $\Gamma \simeq
(\alpha_{2,3}^2/4\pi) (m_{3/2}/M_{d})^2 M_{d}$ and the
resultant decay temperature is $T^{\rm mess}_d \simeq 10$ GeV
$\sqrt{(f/10^{-3})(k_d/10^{-1})}(m_{3/2}/{\rm MeV})$.    
As discussed in the next section, the relic abundance of the messenger
quarks and leptons are give by Eq. (\ref{eq:messrelic}) after the
reheating of the thermal inflation. 
Then the temperature at which the messenger quarks and leptons could
begin to dominate the energy density if they were stable is given by 
\begin{eqnarray}
 T^{\rm mess}_c \simeq M_{d,l}\frac{4}{3}
\bigg(\frac{n^{\rm after}_{d,l}}{s(T^{\rm mess}_c)}\bigg) 
\simeq 10^{-4} {\rm GeV} \bigg(\frac{M_{d,l}}{10^7 \GEV}\bigg)^3 
\bigg(\frac{10^5 \GEV}{m_S}\bigg)\bigg(\frac{f}{10^{-3}}\bigg),
\label{eq:messenger}
\end{eqnarray}
where $n^{\rm after}_{d,l}$ are the number density of the messenger
quarks and leptons (see Eq. (\ref{eq:messrelic})).
We see that $T^{\rm mess}_c$ is much smaller than $T^{\rm mess}_d$,
and hence the messenger quarks  and leptons produce no significant
entropy in the present scenario.   

We conclude that none of quasi-stable particles (except for the $R$
axion) in the DSB and the messenger sectors produces extra entropy after
the freeze-out time of the gravitino. 

\begin{figure}[htb]
\begin{picture}(100,80)(20,0)
 \ArrowLine(10,50)(50,50)
 \Vertex(50,50){2}
 \ArrowLine(50,50)(80,70)
 \ArrowLine(80,30)(50,50)
 \Text(20,40)[]{$Z^{ij}$}
 \Text(90,70)[]{$E^m$}
 \Text(90,30)[]{$E^{l\dagger}$}
 \Text(45,65)[]{\footnotesize{$\displaystyle{\big[\frac{M_E}{M_G}}\big]$}}
\end{picture}
\begin{picture}(100,100)(0,0)
 \ArrowLine(25,50)(-8,50)
 \ArrowLine(25,50)(50,50)
 \ArrowLine(50,50)(80,70)
 \ArrowLine(80,30)(50,50)
 \Vertex(50,50){2}
 \Line(22,47)(28,53)
 \Line(28,47)(22,53)
 \Text(40,40)[]{$Z^{ij}$}
 \Text(0,40)[]{$(QQ)$}
 \Text(23,60)[]{$M_{Z}$}
 \Text(90,70)[]{$E^m$}
 \Text(90,30)[]{$E^{l\dagger}$}
 \Text(50,70)[]{\footnotesize{$\displaystyle{\big[\frac{M_E}{M_G}}\big]$}}
\end{picture}
\hspace{.5cm}
\begin{picture}(100,100)(0,15)
 \ArrowLine(0,70)(40,70)
 \ArrowLine(40,70)(80,70)
 \ArrowLine(0,65)(40,65)
 \ArrowLine(40,65)(80,65)
 \ArrowLine(40,65)(60,45)
 \Vertex(40,65){2}
 \Text(70,45)[]{$E^i$}
 \Text(90,75)[]{$Q'$}
 \Text(88,63)[]{$Q$}
 \Text(-7,75)[]{$Q'$}
 \Text(-7,63)[]{$Q'$}
 \Text(35,50)[]{\footnotesize{$\displaystyle{\big[\frac{\Lambda_H}{M_G}}\big]$}} 
\end{picture}
\hspace{.5cm}
\vspace{.6cm}
\begin{picture}(100,100)(0,15)
 \ArrowLine(0,70)(40,70)
 \ArrowLine(40,70)(80,70)
 \ArrowLine(0,65)(40,65)
 \ArrowLine(40,65)(80,65)
 \ArrowLine(40,65)(60,45)
 \Vertex(40,65){2}
 \Text(70,45)[]{$E^i$}
 \Text(90,75)[]{$Q$}
 \Text(90,63)[]{$Q$}
 \Text(-7,75)[]{$Q'$}
 \Text(-9,63)[]{$Q$}
 \Text(35,50)[]{\footnotesize{$\displaystyle{\big[\frac{\Lambda_H}{M_G}}\big]$}} 
\end{picture}
\vspace{-1.5cm}
\caption{\footnotesize{The super Feymnan diagrams for decay processes of
 $Z^{ij}$, $(QQ)$, $(Q'Q')$ and $(QQ')$.}}
\label{fig:DSBdecay}
\begin{center}
 \begin{picture}(200,70)(0,0)
  \ArrowLine(130,60)(80,60)
  \ArrowLine(130,30)(80,30)
  \ArrowLine(30,60)(80,60)
  \ArrowLine(30,30)(80,30)
  \ArrowLine(80,45)(80,60)
  \ArrowLine(80,45)(80,30)
  \Line(78,43)(82,47)
  \Line(78,47)(82,43)
  \Text(140,60)[]{$S$}
  \Text(140,30)[]{$S$}
  \Text(10,60)[]{$E^i$}
  \Text(10,30)[]{$E^j$}
  \Text(105,45)[]{$k_{ij}\vev{S}$}
\end{picture}
\hspace{.5cm}
 \begin{picture}(200,70)(0,0)
\Photon(20,60)(60,60){2}{5}
\Photon(20,30)(60,30){2}{5}
\ArrowLine(140,60)(100,60)
\ArrowLine(100,30)(140,30)
\ArrowLine(60,60)(100,60)
\ArrowLine(100,30)(100,60)
\ArrowLine(100,30)(60,30)
\ArrowLine(60,30)(60,60)
\Vertex(100,30){2}
\Vertex(100,60){2}
\Text(10,45)[]{\footnotesize{SU(2)$_m$ glue ball}}
\Text(80,70)[]{$E_i$}
\Text(150,60)[]{$S^{\dagger}$}
\Text(150,30)[]{$S$}
\Oval(10,45)(12,45)(360)
\end{picture}
\end{center}
\vspace{-1.3cm}
\caption{\footnotesize{The super Feymnan diagrams for decay processes of
the $E^iE^j$ bound states and the SU(2)$m$ glue  balls.
}}
\label{fig:decayE}
 \begin{center}
 \begin{picture}(200,70)(0,0)
  \ArrowLine(20,45)(60,45)
  \ArrowLine(80,65)(60,45)
  \ArrowLine(80,25)(60,45)
  \ArrowLine(80,65)(80,45)
  \ArrowLine(80,25)(80,45)
  \Photon(80,65)(120,65){2}{4}
  \Photon(80,25)(120,25){2}{4}
  \Line(78,43)(82,47)
  \Line(78,47)(82,43)
  \Vertex(60,45){2}
  \Text(10,45)[]{$S$}
  \Text(95,45)[]{$\vev{S}$}
  \Text(70,70)[]{$d_M$}
  \Text(70,20)[]{$\bar{d}_M$}
  \Text(180,65)[]{SSM\mbox{ gauge multiplet}}
  \Text(180,25)[]{SSM\mbox{ gauge multiplet}}
 \end{picture}
 \end{center}
 \vspace{-1.0cm}
 \caption{\footnotesize{The super Feymnan diagrams for decay processes of S.
 }}
 \label{fig:decayS}
\end{figure}

\section{A mini-thermal inflation and the entropy
 production}\label{sec:thermalif}

In this section we discuss the thermal history of the present system. 
We consider the reheating temperature $T_R$ is much higher than the DSB
scale, $\Lambda_H \simeq 10^{8-9}$ GeV, and all particles in the DSB and
the messenger sectors as well as the SSM sector including the gravitino
are in the thermal bath. 
Then, the expectation value of the field $S$ is set at the origin by the 
thermal effects. 
When the temperature $T$ cools down to the messenger scale, the radial
component of the scalar field ${S}$ starts rolling down to the true
minimum from the origin. 
We call it the ``flaton'' $S$.   
From Eq. (\ref{eq:effpotential}) it is clear that if the coupling 
constant $f$ is small, the potential of $S$ is very flat and a thermal
inflation \cite{Lyth:1995ka} takes place.  
We assume, for the time being, that it is the case and calculate how much
the entropy is produced after the thermal inflation. 
And we show, later on, that the coupling $f$ is naturally small as
$f\simeq 10^{-2} - 10^{-4}$, while other Yukawa coupling
constants, $k_E$, $k_{d}$, $k_{l}={\cal O}(1)$.     

When the temperature $T$ reaches $T \simeq m_S /(2\sqrt{f})$, the energy
density of the field ${S}$, $\rho_{\rm start} \simeq m_S^4/(4f^2)$,  
begins to dominate the total energy density of the universe and the
thermal inflation starts.   
It ends when the temperature falls down to $T\simeq m_S$, and
the ``flaton'' $S$ starts to oscillate around the minimum of the
potential. 
The ``flaton'' $S$ decays into $R$ axions dominantly as explained in the
previous section. 
Thus, the decay of ``flaton'' $S$ only reheats up the temperature of the
$R$ axion, while the temperature of the SSM sector unreheated
\cite{Asaka:1999xd}.  
Although the decay of the ``flaton'' $S$ occurs sufficiently fast in the
vacuum,  the reheating temperature  $T_{\rm th}$ of $R$ axion cannot
exceed the mass of the ``flaton'' $S$, and hence the reheating
temperature $T_{\rm th}$ is fixed by the mass of the ``flaton'' $S$
field, that is  $T_{\rm th}$ $\simeq \sqrt{2}m_S$ \cite{Kolb:2003ke}.    
The resultant yield of the gravitino is given by 
\begin{eqnarray}
 Y^{\rm after}_{3/2} \equiv \frac{n^{\rm after}_{3/2}}{s^{\rm after}} 
= \frac{1}{s^{\rm after}}
\bigg(\frac{\rho^{\rm after}}{\rho^{\rm before}}\bigg){n^{\rm before}_{3/2}} 
= \frac{3}{4}T_{\rm th} \bigg(\frac{s^{\rm before}}{\rho^{\rm before}}\bigg)
\frac{n^{\rm before}_{3/2}}{s^{\rm before}}
\simeq \bigg(\frac{\pi^2g^{\rm before}_*}{30}\bigg)(4\sqrt{2}f^2)Y^{\rm before}_{3/2},
\label{eq:dilution1}
\end{eqnarray}
where $Y^{\rm after,before}_{3/2}$ are the yields of the gravitino
after/before the decay of the ``flaton'' $S$, $n^{\rm
after,before}_{3/2}$ the number densities of the gravitino, $\rho^{\rm
after, before}$ the energy densities and $s^{\rm after, before}$ the
entropy densities.\footnote{ 
The $g^{\rm before}_*\simeq 230$ is the degree of freedoms of effective
massless particles just after the end of the thermal inflation, which
corresponds to the number of the SSM particles.
On the other hand, we use $g^{\rm after}_* \simeq 1$ for the degree of
freedoms of massless particles in the $R$ axion (radiation) dominant
era. 
}  
Here, we have used the fact that the universe is matter dominant during
the decay of the ``flaton'' $S$, $\rho^{\rm after}/s^{\rm after}=
(3/4)T_{\rm th}$, $s^{\rm before}\simeq (2 \pi^2g^{\rm before}_*/45)
m_S^3$ and $\rho^{\rm before} \simeq m_S^4/4f^2$.
After the decay of the ``flaton'' $S$, the $R$ axions dominate the
energy 
density of the universe and when the temperature of the $R$ axion cools
down to its decay temperature $T^{\rm axion}_d$, they decay into pairs of
SM gluons.  
Then, the SM particles are reheated up and the yield of the gravitino is
further diluted as 
\begin{eqnarray}
  Y^{\rm af.decay}_{3/2} 
= \frac{1}{s^{\rm af.decay}}
\bigg(\frac{\rho^{\rm af.decay}}{\rho^{\rm bef.decay}}\bigg)
{n^{\rm bef.decay}_{3/2}}  
= \frac{3}{4}T^{\rm axion}_d 
\bigg(\frac{s^{\rm bef.decay}}{\rho^{\rm bef.decay}}\bigg) 
Y^{\rm bef.decay}_{3/2}
\simeq\frac{3}{4}\frac{T^{\rm axion}_d}{m_{\rm axion}}Y^{\rm bef.decay}_{3/2},
\label{eq:dilution2}
\end{eqnarray}
where superscript af.decay/bef.decay means the after/before the $R$
axion decay.
Here, we have used that the universe is $R$ axion-matter dominant when
the $R$ axion decays and also used $\rho^{\rm bef.decay} = s^{\rm
bef.decay} m_{\rm axion}$.  
From the Eq. (\ref{eq:dilution1}) and Eq. (\ref{eq:dilution2}), we
obtain the resultant dilution factor of the gravitino as
\begin{eqnarray}
\Delta \equiv \bigg(\frac{Y^{\rm before}_{3/2}}{Y^{\rm af.decay}_{3/2}}\bigg) 
 =\bigg(\frac{Y^{\rm before}_{3/2}}{Y^{\rm after}_{3/2}}\bigg) 
\bigg(\frac{Y^{\rm bef.decay}_{3/2}}{Y^{\rm af.decay}_{3/2}}\bigg) 
 \simeq \frac{4}{3}\frac{m_{\rm axion}}{T^{\rm axion}_d}
\bigg(\frac{30}{\pi^2 g^{\rm before}_{*}}\bigg) \frac{1}{4\sqrt{2} f^2}, 
\label{eq:thermal}
\end{eqnarray}
where we have used $Y^{\rm bef.decay}_{3/2}=Y^{\rm after}_{3/2}$, since
no extra entropy for the SSM particles is produced when the ``flaton''
$S$ decays  and the yield of the gravitino does not change in the $R$
axion dominant era.    

Here, we estimate the relic abundances of the messenger quarks and
leptons.
When the ``flaton'' $S$ stay at the origin, the messenger quarks and leptons
are massless, and their annihilation processes take place during the
reheating epoch of the thermal inflation.
When the annihilation rates $\vev{\sigma v}n_{d,l}$ become smaller than the
Hubble expansion rate $H$, the messenger quarks and leptons are frozen 
out from the thermal bath with the number density $n^f_{d,l} \simeq
H_f/\vev{\sigma v}$, where $\vev{\sigma v}$ is a annihilation cross
section of the messenger quarks and leptons and the sub(super)script $f$ 
denotes the ``freeze-out'' time.  
Since the annihilation processes are instantaneous, $H_f$ is estimated
as $H_f \simeq \sqrt{\rho^{\rm before}}/M_G$. 
Thus, the resultant relic abundances of the messenger quarks and leptons
after the reheating process are given by
\begin{eqnarray}
 \frac{n_{d,l}^{\rm after}}{s^{\rm after}} \simeq \frac{1}{s^{\rm after}}
\bigg(\frac{\rho^{\rm after}}{\rho^{\rm before}}\bigg) n^f_{d,l}
\simeq \frac{2f}{\vev{\sigma v} M_G m_S}
\simeq \frac{2f M_{d,l}^2}{4\pi\alpha_{2,3}^2 M_G m_S},
 \label{eq:messrelic}
\end{eqnarray}
where we have used the fact that the universe is matter dominant during
the decay of the ``flaton'' $S$, $\rho^{\rm after}/s^{\rm after}=
(3/4)T_{\rm th}$, $\rho^{\rm before} \simeq m_S^4/4f^2$ and $\vev{\sigma
v}\simeq 4 \pi \alpha_{2,3}^2/M_{d,l}^2$.  
As discussed in the previous section, the messenger quarks and
leptons can not dominate the energy density of the universe before their
decay times, and hence they produce no extra entropy.\footnote{ 
The bound states of the hyperquarks are heavy even  before the thermal
inflation. 
Thus, their annihilation processes finish before the thermal
inflation. 
Therefore, their relic abundances are diluted by the thermal inflation
as well as the relic abundance of the gravitino (Eq. (\ref{eq:Z})).}  

Now, we estimate the dilution factor $\Delta$ needed to explain the
mass density of the dark matter by the stable gravitinos. 
If there were no entropy production the yield of the thermal gravitinos
could be given by 
\begin{eqnarray}
  Y_{3/2} \equiv \frac{n_{3/2}}{s} \simeq \frac{45}{2\pi^2g_*(T_f)}
\frac{\zeta (3)}{\pi^2}
\bigg(\frac{3}{2}\bigg),
\label{eq:gravitinoyield}
\end{eqnarray}
where $n_{3/2}$ is the number density of gravitinos and $T_f$ is the
freeze-out temperature of the gravitinos.\footnote{
In the present scenario, the temperature where the gravitinos are
thermalized is higher than the reheating temperature $T_{\rm th}$ of the
thermal inflation.
}
In terms of the density parameter it is\footnote{
At higher temperatures than the reheating temperature $T_{\rm th}$, the
expectation value of the field $S$ is set at the origin, and the fields
of the messenger sector are also massless. 
Therefore, the degree of freedoms of the effective massless particles, 
$g_*(T_f)$, is enhanced as $g_*(T_f) \simeq 350$ for $T_f > T_{\rm
th}$. 
} 
\begin{eqnarray}
  \Omega_{3/2}h^2 \simeq 5.0 \times 10^2 \bigg(\frac{m_{3/2}}{1 \rm MeV}\bigg)
\bigg(\frac{350}{g_*(T_f)}\bigg),
\label{eq:gravitinodensity}
\end{eqnarray}
where $h$ is the present Hubble parameter in units of 100 km sec$^{-1}$
Mpc$^{-1}$, and $\Omega_{3/2} = \rho_{3/2}/\rho_c$. 
Here, $\rho _{3/2}$ and $\rho_c$ are the energy density of the gravitino
and the critical density in the present universe, respectively. 
The required dilution factor $\Delta$ is given by
\begin{eqnarray}
  \Delta_{r} \simeq  3.0 \times 10^3 \bigg(\frac{m_{3/2}}{1 \rm MeV}\bigg)
\bigg(\frac{350}{g_*(T_f)}\bigg)\bigg(\frac{0.11}{\Omega_{\rm DM}h^2}\bigg),
\label{eq:req}
\end{eqnarray}
to account for the observation of the dark matter density
$\Omega_{DM}h^2 \simeq 0.11$.  
From Eqs. (\ref{eq:thermal}) and (\ref{eq:req}) we see $f\simeq
10^{-2.8}$ for $m_{3/2} = O(1)$ MeV and $m_{\rm axion}\simeq 10$ GeV.

In the followings, we show that the coupling constant $f$ is naturally
small at the messenger scale $m_S$.
We assume that all Yukawa coupling constants in the messenger sector is
of order of unity at the gravitational scale $M_G$ and $n_q =
3$.\footnote{
For $n_q\ge 4$ the messenger gauge coupling constant $\alpha_m$ is
non asymptotic free.
}
All Yukawa coupling constants including $f$ at the messenger scale $m_S$
are determined by solving the following renormalization group equations 
(RGEs); 
\begin{eqnarray}
 \frac{\partial}{\partial\ln\mu}\bigg(\frac{1}{\alpha_f}\bigg) =
-\bigg[
\frac{6}{2\pi}
+ \frac{6}{2\pi}\bigg(\frac{\alpha_l}{\alpha_f}\bigg)
+ \frac{9}{2\pi}\bigg(\frac{\alpha_d}{\alpha_f}\bigg)
+ \frac{24}{2\pi}\sum_{i>j}\bigg(\frac{\alpha_E^{ij}}{\alpha_f}\bigg)
\bigg],
\label{eq:RGEf}
\end{eqnarray}
\begin{eqnarray}
 \frac{\partial}{\partial\ln\mu}\bigg(\frac{1}{\alpha_d}\bigg) =
-\bigg[
\frac{5}{2\pi} 
+ \frac{2}{2\pi}\bigg(\frac{\alpha_l}{\alpha_d}\bigg)
&+& \frac{2}{2\pi}\bigg(\frac{\alpha_f}{\alpha_d}\bigg)
+ \frac{8}{2\pi}\sum_{i>j}\bigg(\frac{\alpha_E^{ij}}{\alpha_l}\bigg)\\\nonumber
&-& \frac{2}{15\pi}\bigg(\frac{\alpha_1}{\alpha_d}\bigg)
- \frac{8}{3\pi}\bigg(\frac{\alpha_3}{\alpha_d}\bigg)
\bigg],
\label{eq:RGEd}
\end{eqnarray}
\begin{eqnarray}
 \frac{\partial}{\partial\ln\mu}\bigg(\frac{1}{\alpha_l}\bigg) =
-\bigg[
\frac{4}{2\pi} 
+ \frac{3}{2\pi}\bigg(\frac{\alpha_d}{\alpha_l}\bigg)
&+& \frac{2}{2\pi}\bigg(\frac{\alpha_f}{\alpha_l}\bigg)
+ \frac{8}{2\pi}\sum_{i>j}\bigg(\frac{\alpha_E^{ij}}{\alpha_l}\bigg)\\\nonumber
&-& \frac{3}{10\pi}\bigg(\frac{\alpha_1}{\alpha_l}\bigg)
- \frac{3}{2\pi}\bigg(\frac{\alpha_3}{\alpha_l}\bigg)
\bigg],
\label{eq:RGEl}
\end{eqnarray}
\begin{eqnarray}
 \frac{\partial}{\partial\ln\mu}\bigg(\frac{1}{\alpha_E^{ij}}\bigg) =
-\bigg(\frac{1}{\alpha_E^{ij}}\bigg)
\bigg[
\frac{2}{2\pi}\alpha_f 
+ \frac{3}{2\pi}\alpha_d
+ \frac{2}{2\pi}\alpha_l
+ \frac{8}{2\pi}\sum_{i>j}\alpha_E^{ij}\\\nonumber
+ \frac{4}{2\pi}\sum_{\ell\neq i}^{6}\alpha_E^{i\ell}
+ \frac{4}{2\pi}\sum_{\ell\neq j}^{6}\alpha_E^{j\ell}
-\frac{3}{2\pi} \alpha_m\bigg]\\\nonumber
+ 
\frac{4}{2\pi}\sum_{\ell\neq i,j}^{6}
\bigg(\frac{1}{\alpha_E^{ij}}\bigg)^{3/2}
\bigg[
\sum_{m\neq \ell,i}^6
\sqrt{\alpha_E^{j\ell}\alpha_E^{\ell m}\alpha_E^{mi}}
+ \sum_{m\neq \ell,j}^6
\sqrt{\alpha_E^{i\ell}\alpha_E^{\ell m}\alpha_E^{mj}}
\bigg],
\label{eq:RGEkE}
\end{eqnarray}
where $\alpha_f = f^2/4\pi$, $\alpha_d = k_d^2/4\pi$, $\alpha_l =
k_l^2/4\pi$ and $\alpha_E^{ij} = k_{ij}^2/4\pi$. 
We can see that the RGE of the coupling constant $f$ has no
effect from the gauge coupling constants which slacken the speed of 
the Yukawa-coupling running.
Thus,  we can expect that the coupling constant $f$ becomes much smaller
than the other Yukawa coupling constants at the low energy scale.

\begin{figure}[htb]
\begin{center}
\begin{minipage}{0.48\linewidth}
\begin{center}
\vspace{-.3cm}
\hspace{-.1cm}
\includegraphics[width=0.78\linewidth]{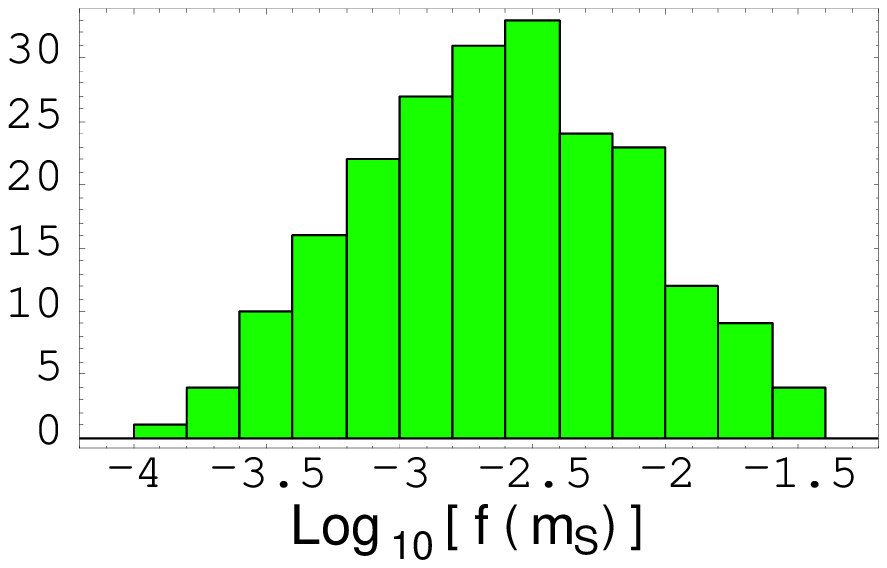} 
\end{center}
\end{minipage}
\begin{minipage}{0.48\linewidth}
\begin{center}
\hspace{-.8cm}
\includegraphics[width=0.82\linewidth]{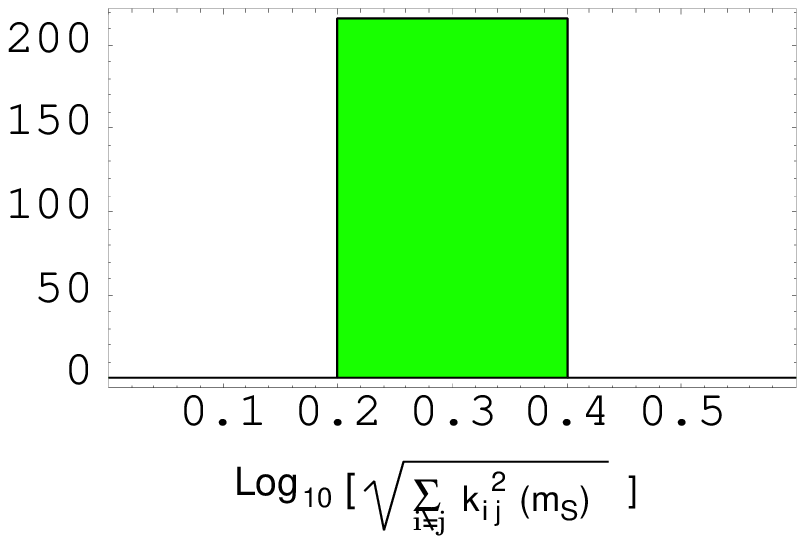} 
\end{center}
\end{minipage}
\begin{minipage}{0.48\linewidth}
\begin{center}
\includegraphics[width=0.78\linewidth]{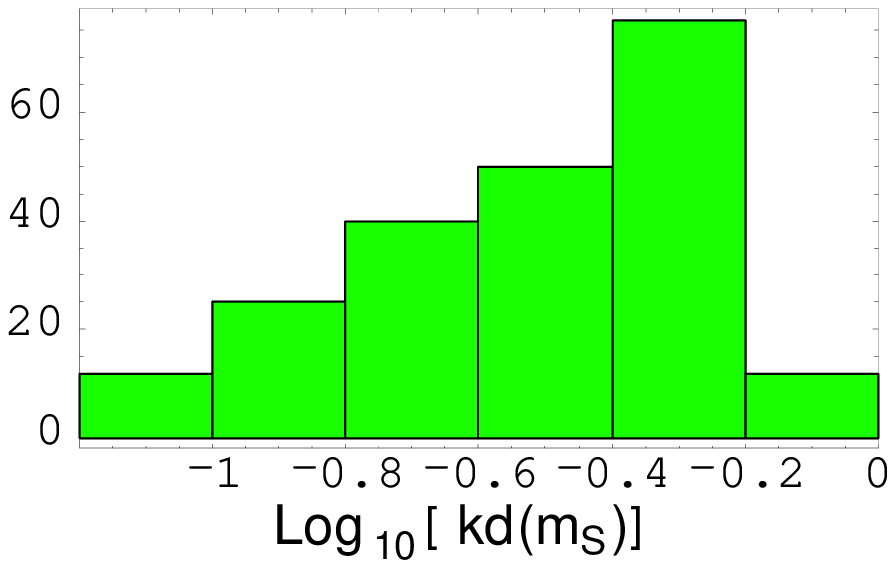} 
\end{center}
\end{minipage}
\begin{minipage}{0.48\linewidth}
\begin{center}
\includegraphics[width=0.78\linewidth]{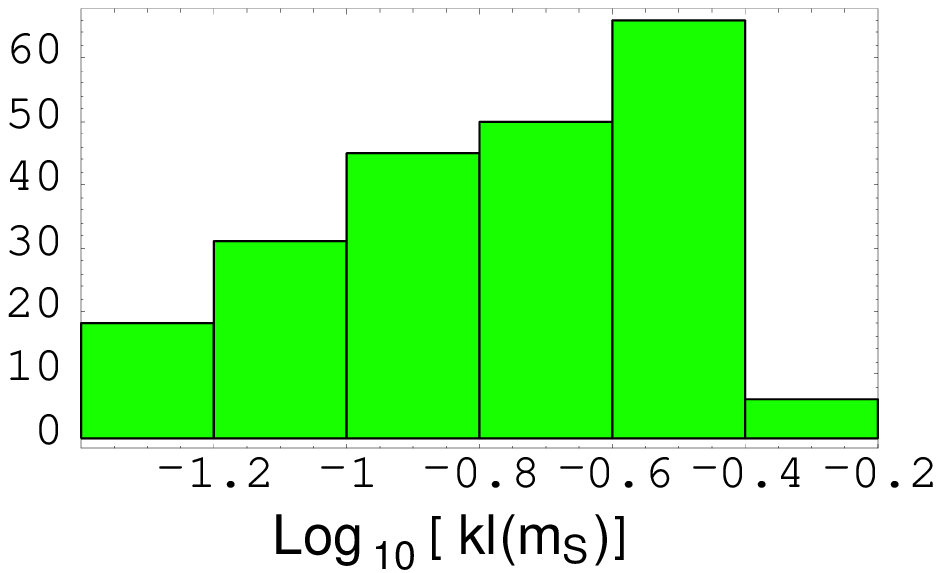} 
\end{center}
\end{minipage}
\caption{
The histograms for Yukawa coupling constants at the DSB scale 
with $n_q = 3$.
We have varied the coupling constants $k_{ij}$, $k_d$, $k_l$ and $f$
 from  0.3 to 3 with a logarithmic measure at the gravitational scale
 $M_G$. 
Here, $\alpha_{m}(\Lambda_H) = 0.5$, and we have set the all  $k_{ij}$
 equal, for simplicity.} 
\label{fig:Histogram}
\end{center}
\end{figure}

The result on the coupling $f$ is shown in Fig. \ref{fig:Histogram}.
Here, we have assumed  
\begin{eqnarray}
  k_{ij},~~k_d,~~k_\ell,~~f =0.3-3,
\end{eqnarray}
at the gravitational scale $M_G$. 
We find that the desired coupling $f\simeq 10^{-3}$ is obtained at the
messenger scale $\mu = m_S$.  
We also show the obtained coupling constants for $\sqrt{\sum_{i\neq
j}k_{ij}}$, $k_d$, $k_\ell$ in Fig. \ref{fig:Histogram}. 
We see that the assumptions on the Yukawa coupling constants made in the
previous section are realized naturally (for instance,
$\sqrt{\sum_{i\neq j}k_{ij}} = {\cal O}(1)$ and $k_d$, $k_l \gg
f$).

Before closing this section, we should comment on the reproductions of 
gravitinos from the thermal background after the thermal inflation.   
We find that the gravitino reproduction rate from the thermal $R$ axion
bath is small enough not to spoil the successful dilution of the
gravitino.  
The reheating temperature of the SSM sector is $T^{\rm axion}_d\simeq 5
{\rm GeV}({m_{\rm axion}}/10 {\rm GeV})$ and hence the gravitino
reproduction from the SSM background is also negligible.  

\section{Conclusions}

\begin{figure}[ht]
\begin{center}
\includegraphics[width=0.5\linewidth]{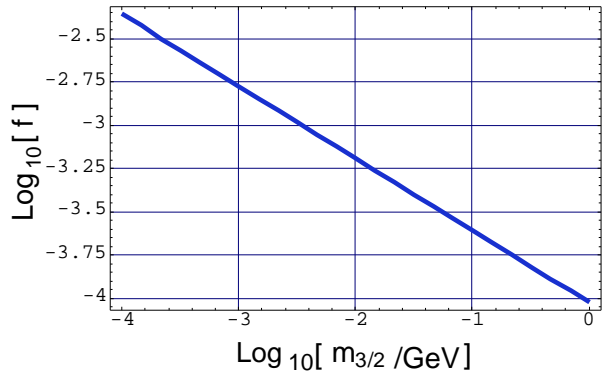} 
\caption{The required coupling constant $f$ which leads to the
 sufficient thermal inflation making the gravitino density
 $\Omega_{3/2}h^2 \simeq 0.11$.}  
\label{fig:neededf}
\end{center}
\end{figure}

We assume, in this paper, that the reheating temperature $T_R$ of the 
primary inflation is $T_R > 10^{10}$ GeV so that the thermal
leptogenesis takes place. 
With this reheating temperature the gravitinos of mass $m_{3/2} \lsim 1$
GeV are thermally produced and they overclose the universe if there is
no entropy production after their freeze-out time.
We find, however, that a mini-thermal inflation occurs naturally in a
class of gauge meditation models we discuss in this paper. 
The reheating process of the thermal inflation produces an amount of
entropy, which dilutes the number density of the relic gravitinos
avoiding the overclosure. 
This dilution makes the gravitino to be the dark matter in the present
universe. 
The dilution factor depends on a Yukawa coupling constant $f$. 
Fig. \ref{fig:neededf} shows the coupling constant $f$ versus the
gravitino mass required to realize the gravitino dark matter,
$\Omega_{3/2}h^2 \simeq 0.11$.  
We see that for $m_{3/2} = 100$ keV $-1$ GeV we need $f\simeq
10^{-2}-10^{-4}$, which is naturally obtained in the present gauge
mediation model. 

The abundance of the gravitino dark matter is independent of the
reheating temperature $T_R$ of the primary inflation, once the
gravitinos are in the thermal equilibrium. 
Therefore, there is no upper bound on the reheating temperature $T_R$
from the overproduction of gravitinos and hence the thermal leptogenesis
takes place without any gravitino problem \cite{Fujii:2002fv}. 
The temperature $T_L$ of the leptogenesis is found as $T_L \simeq
10^{12-14}$ GeV for $m_{3/2} \simeq 100$ keV$-10$ MeV (see
Ref.\cite{Fujii:2002fv}). 

\section*{Acknowledgment}

The authors wish to thank M.Kawasaki for a useful discussion.
M.F. would like to thank the Japan Society for the Promotion of Science
for financial support.
This work is partially supported by Grand-in-Aid Scientific Research (s)
14102004. 



\begin{thebibliography}{99}
%
\bibitem{Fukugita:1986hr}
M.~Fukugita and T.~Yanagida,
Phys.\ Lett.\ B {\bf 174}, 45 (1986).


%
\bibitem{BBP}
W.~Buchmuller, P.~Di Bari and M.~Plumacher,
Nucl.\ Phys.\ B {\bf 665}, 445 (2003).
%
\bibitem{Weinberg:zq}
S.~Weinberg,
Phys.\ Rev.\ Lett.\  {\bf 48}, 1303 (1982).
\bibitem{Holtmann:1998gd}
E.~Holtmann, M.~Kawasaki, K.~Kohri and T.~Moroi,
Phys.\ Rev.\ D {\bf 60}, 023506 (1999);
M.~Kawasaki, K.~Kohri and T.~Moroi,
Phys.\ Rev.\ D {\bf 63}, 103502 (2001).
%
\bibitem{Kohri:2001jx}
K.~Kohri,
Phys.\ Rev.\ D {\bf 64}, 043515 (2001);
K.~Kohri, talk given at  COSMO-2003;
M.~Kawasaki, private communication.
%


\bibitem{gauge-mediation}
M.~Dine and A.~E.~Nelson,
Phys.\ Rev.\ D {\bf 48}, 1277 (1993);
M.~Dine, A.~E.~Nelson and Y.~Shirman,
Phys.\ Rev.\ D {\bf 51}, 1362 (1995);
M.~Dine, A.~E.~Nelson, Y.~Nir and Y.~Shirman,
Phys.\ Rev.\ D {\bf 53}, 2658 (1996);
For a review, see, for example, 
G.~F.~Giudice and R.~Rattazzi,
Phys.\ Rept.\  {\bf 322}, 419 (1999).
%
\bibitem{Luty:1997fk}
M.~A.~Luty,
Phys.\ Rev.\ D {\bf 57}, 1531 (1998).
\bibitem{Moroi:1993mb}
T.~Moroi, H.~Murayama and M.~Yamaguchi,
Phys.\ Lett.\ B {\bf 303}, 289 (1993).
\bibitem{Fujii:2002fv}
M.~Fujii and T.~Yanagida,
Phys.\ Lett.\ B {\bf 549}, 273 (2002).
%
\bibitem{Nomura:1997ur}
Y.~Nomura, K.~Tobe and T.~Yanagida,
Phys.\ Lett.\ B {\bf 425}, 107 (1998).
%
\bibitem{Izawa:1996pk}
K.~I.~Izawa and T.~Yanagida,
Prog.\ Theor.\ Phys.\  {\bf 95}, 829 (1996).
\bibitem{Intriligator:1996pu}
K.~A.~Intriligator and S.~Thomas,
Nucl.\ Phys.\ B {\bf 473}, 121 (1996).
%
\bibitem{Vilenkin:zs}
A.~Vilenkin,
Phys.\ Rev.\ D {\bf 23}, 852 (1981).
\bibitem{Izawa:1997hu}
K.~I.~Izawa,
Prog.\ Theor.\ Phys.\  {\bf 98}, 443 (1997).


%
\bibitem{Lyth:1995ka}
D.~H.~Lyth and E.~D.~Stewart,
Phys.\ Rev.\ D {\bf 53}, 1784 (1996).
%
\bibitem{Asaka:1999xd}
T.~Asaka and M.~Kawasaki,
Phys.\ Rev.\ D {\bf 60}, 123509 (1999).
\bibitem{Kolb:2003ke}
E.~W.~Kolb, A.~Notari and A.~Riotto,
arXiv:hep-ph/0307241.
%


\end{thebibliography}
\end{document}